\documentclass[11pt]{article}
\usepackage{amsmath}
\usepackage[T1]{fontenc}
\usepackage{charter}
\usepackage{microtype}

\usepackage{geometry}
\def\0{\boldsymbol{0}}

\def\eL{\mathbf L}
\def\eps{\varepsilon}

\def\R{\mathbf R}

\def\H{\mathcal H}

\newcommand{\w}{\omega}
\newcommand{\inner}[1]{\left\langle #1\right\rangle}
\newcommand{\abs}[1]{\left\lvert #1\right\rvert}

\newenvironment{remark}[1][Remark:]{\begin{trivlist}
\item[\hskip \labelsep {\bfseries #1}]}{\end{trivlist}}

\title{$\eL^p$-boundedness of the Hilbert transform}

\author{Kunal Narayan Chaudhury\thanks{kchaudhu@math.princeton.edu. } }

\begin{document}

\date{}

\maketitle

\begin{abstract} The Hilbert transform is essentially the \textit{only} singular operator in one dimension. This undoubtedly makes it one of the the most important linear operators in harmonic analysis. The Hilbert transform has had a profound bearing on several theoretical and physical problems across a wide range of disciplines; this includes problems in Fourier convergence, complex analysis, potential theory, modulation theory, wavelet theory, aerofoil design, dispersion relations and high-energy physics, to name a few.

	In this note, we revisit some of the established results concerning the global behavior of the Hilbert transform, namely that it is is weakly bounded on $\eL^1(\R)$, and strongly bounded on $\eL^p(\R)$ for $1 < p  <\infty$, and provide a self-contained derivation of the same using real-variable techniques. This note is partly based on the expositions on the Hilbert transform in \cite{Grafakos, Stein}.
\end{abstract}

\section{Introduction}

The Hilbert transform of a sufficiently well-behaved function $f(x)$ is defined to be
\begin{equation}
\label{def}
\H f(x)=\frac{1}{\pi} \lim_{\varepsilon \rightarrow 0} \int_{|t| > \varepsilon} f(x-t) \frac{dt}{t}.
\end{equation}
It is immediately not clear that $\H f(x)$ is well-defined even for nice functions. Though \eqref{def} ``almost'' looks like an ordinary convolution, there are however certain technical subtleties associated with the definition. The primitive idea behind the definition of the transform is quite simple, namely to transform $f(x)$ by convolving with the kernel $1/\pi x$. It is in doing so rigorously that one encounters technical difficulties -- the kernel fails to be absolutely integrable owing to its slow decay and, more importantly, due to the singularity at the origin. The  limiting argument in \eqref{def} is used to avoid the singularity by truncating the kernel around the origin in a systematic fashion. As will be shown shortly, this indeed works for sufficiently regular functions. The other pathology, namely the slow decay of the kernel, can be be circumvented relatively easily simply by restricting the domain of \eqref{def} to functions having a sufficiently fast decay.

The non-trivial task in the study of the Hilbert transform is, in fact, the specification of the class of functions on which the sequence of integrals in \eqref{def} can be given a precise meaning, either pointwise or in the norm sense. More precisely, one needs to show the integral 
\begin{equation*}
\H_{\varepsilon} f(x)=\frac{1}{\pi}  \int_{|t| > \varepsilon} f(x-t) \frac{dt}{t}
\end{equation*}
is (absolutely) convergent  for all $\varepsilon >0$ and that (i) either $\H_{\varepsilon} f(x)$ converges for (almost) all $x$ as $\varepsilon \rightarrow 0$, which provides a pointwise specification of $\H f$; or (ii) that the sequence of functions $\H_{\varepsilon} f$ converge in the norm to some function in $\eL^p$ as $\varepsilon \rightarrow 0$, which is then defined to be the Hilbert transform of $f$.

	Here we will focus only on the latter global characterization of $\H$, namely the fact that it is weakly bounded on $\eL^1$ (marginally fails to be bounded), and that it is strongly bounded on $\eL^p=\eL^p(\R)$ for $1 < p  <\infty$; the latter result was originally derived by M. Riesz using techniques from complex analysis \cite{Riesz}. We will however use real-variable techniques (see \cite{Stein}) and will particularly focus on the main ideas rather than the technical details.

\section{Details of the derivation}

	To begin with, we restrict the domain of $\H$ to the Schwartz class $\mathbf{S}=\mathbf{S}(\R)$ on which $\H f_{\varepsilon}(x)$ is well-defined for all $x$ and for every $\varepsilon >0$. Indeed, following the fact that the $p$-th power of $1/|t|$ is integrable outside the interval $(-\varepsilon,\varepsilon)$ for all $1 <p < \infty$, Holder's inequality tells us that $\H_{\varepsilon} f(x)$ exists for all $\varepsilon >0$. As far as the convergence is concerned, all we need to show is that the integral remains absolutely convergent even as $\eps \longrightarrow 0$. 	To this end, we split the integral in \eqref{def}, and use the anti-symmetric nature of $1/t$ to write
\begin{align*}
\H f(x)  & = \frac{1}{\pi} \lim_{\varepsilon \rightarrow 0} \int_{\eps < |t| < 1} f(x-t) \frac{dt}{t} + \frac{1}{\pi} \int_{|t| \geq 1} f(x-t) \frac{dt}{t} \\
& = \frac{1}{\pi} \lim_{\varepsilon \rightarrow 0} \int_{\varepsilon < |t| < 1}  \frac{f(x-t)-f(x)}{t} dt +\frac{1}{\pi} \int_{|t| \geq 1} f(x-t) \frac{dt}{t}.
\end{align*}
Now, by the mean value theorem, $f(x-t)-f(x)=- t f'(\theta_t)$, where $\theta_t$ is some number between $x$ and $x-t$. Since $f'(x)$ is bounded, we conclude that
\begin{equation*}
 \lim_{\varepsilon \rightarrow 0} \int_{\varepsilon < |t| < 1} \abs{\frac{f(x-t)-f(x)}{t}} dt  \leq  2 \lVert f' \lVert_{\infty}.
\end{equation*}
Thus, $\H f(x)$ is indeed well-defined for all $x$. 

	As a by product of the above observations, we note that the the Hilbert transform of a function can be defined pointwise provided the function exhibits sufficient regularity and decay; in particular, we can modify the above derivation to show that the Hilbert transform of compactly supported function with some Lipschitz regularity is always well-defined.

	Having established the validity of \eqref{def} for the Schwartz class, we next proceed with the derivation of the following  global estimates for this class:
\begin{description}
\item (i) $\H$ is \textit{weak} $(1,1)$:
\begin{equation}
\label{weak11}
m(\{x: |\H f(x)|\geq \lambda\} ) \leq \frac{C}{\lambda} ||f||_1.
\end{equation}
\item (ii) $\H$ is \textit{strong} $(2,2)$:
\begin{equation}
\label{strong22}
\| \H f \|_2 =  \| f \|_2.
\end{equation}
\end{description}
That is, we will show that that $\H$ takes $\eL^1$ into the so-called \textit{weak} $\eL^1$ (a space larger than $\eL^1$), and that it $\H$ maps $\mathbf{S}$ into $\eL^2$ in an isometric manner. Using the fact that $\mathbf{S}$ is dense in $\eL^p$, we can then easily extend these estimates from the sense subclass $\mathbf{S}$ to the larger $\eL^p$ spaces (this is similar to the approximation technique used for extending the domain of the Fourier transform from either $\mathbf{S}$ or $\eL^1$ to $\eL^2$).

	In particular, we will derive the \textit{weak} $(1,1)$ inequality using the decomposition of Calder\'on and Zygmundand, and the \textit{strong} $(2,2)$ inequality using the theory of distributions and the properties of the Fourier transform on $\eL^2$. The \textit{strong} $(p,p)$ boundedness for $1<p\leq 2$, will then be leveraged from these results using a powerful interpolation result. Finally, we will extend the \textit{strong} $(p,p)$ result for $2 <p < \infty$ using duality and the fact that $\H$ is skew-adjoint. 
\begin{remark}
We would like to emphasize on the fact that the foregoing account only tells us that $\H$ is bounded on $\eL^p$ for $1 < p < \infty$ (more precisely, bounded on a dense subclass and thus has a unique extension); it however does not settle the local problem, namely whether \eqref{def} makes sense for $\eL^p$ functions in the notion of pointwise convergence, though it does offers confidence that this indeed must be the case (one needs to study the maximal version of the operator in this case, where the limit in \eqref{def} is replaced by the supremum of the absolute value).
\end{remark}

		We will first derive the estimate in \eqref{strong22} and then the estimate in \eqref{weak11}. We will continue to stress on the main ideas rather than the technical details. 

\subsection{Strong $(2,2)$ nature}

	      Before diving into the details, we would like to note that the bounded nature of $\H$ on $\eL^2$ can be deduced (at least informally) using a argument based on the scaling property of the Fourier transform. Note that since $1/t$ is homogenous of degree $-1$, its Fourier transform $\widehat{(1/t)}$ (provided it indeed is a true function) must necessarily be of degree $0$, that is, it must be bounded. Thus, if we treat \eqref{def} as a convolution between $f(t)$ and the kernel $1/t$, then the convolution-multiplication rule along with the Parseval-Plancherel theorem gives us the estimate
\begin{equation*}
\| \H f \|_2= \| \widehat{\H f} \|_2= \| \widehat{(1/t)} \hat f \|_2 = C \| \hat f \|_2=C \|f \|_2
\end{equation*}
which establishes the fact that $\H$ is bounded on $\eL^2$ (in fact, $C=1$ as will be determined shortly). We will now rigorously show that $\H$ is unitary on $\eL^2$ using the machinery of distribution theory (note this can also be done using classical techniques of complex analysis; e.g., see \cite{Riesz}). 
   
   Consider the distribution $W$ on $\mathbf{S}$, specified by
\begin{equation*}
\inner{W, f} = \frac{1}{\pi} \lim_{\varepsilon \rightarrow 0} \int_{|t| > \varepsilon} f(t) \frac{dt}{t}.
\end{equation*}
It is easily seen that $W$ is linear. Moreover, one can show that the map $f \mapsto \inner{W, f}$ is continuous from $\mathbf{S}$ to $\R$ so that $W$ represents a valid distribution. Given a Schwartz function $f(x)$, note that \eqref{def} can be written as 
\begin{equation*}
\H f(x)=\inner{W, \tau_x f} 
\end{equation*}
where $\tau_x f(t)$ denotes the function $(\tau_x f)(t)=f(x-t)$. 

If we denote the Fourier transform of $W$ by $\widehat W$ and that of $f$ by $\hat f$, then we can use duality to write  
\begin{align*}
\langle \widehat W, f \rangle &= \langle W, \hat f \rangle \\
&=  \frac{1}{\pi} \lim_{\varepsilon \rightarrow 0} \int_{|x| > \varepsilon}^{1/\varepsilon} \hat f(x) \frac{dx}{x} \\
&=  \frac{1}{\pi} \lim_{\varepsilon \rightarrow 0} \int_{|x| > \varepsilon}^{1/\varepsilon}  \Big(\int f(\w) e^{-j \w x } d\w\Big) \frac{dx}{x}.
\end{align*} 
Interchanging the integrals, using Fubini, and applying dominated convergence, we have
\begin{align*} 
\langle \widehat W, f \rangle & = \frac{1}{\pi} \lim_{\varepsilon \rightarrow 0} \int f(\w) \Big(\int_{|x| > \varepsilon}^{1/\varepsilon} \frac{e^{-j \w x }}{x} dx\Big) d\w  \\
& = -\frac{2j}{\pi} \int f(\w) \Big( \lim_{\varepsilon \rightarrow 0} \int_{|x| > \varepsilon}^{1/\varepsilon} \frac{\sin{\w x }}{x} dx\Big) d\w \\
& = -\frac{2j}{\pi}  \int f(\w) \ \frac{\pi}{2}\mathrm{sign}(\w) d\w \\
& =  \int  \left( - j \mathrm{sign}(\w) \right) f(\w) d\w.
\end{align*} 
This tells us that the Fourier transform of $W$ is in fact a function, and is given by 
\begin{equation*}
\widehat W(\w)=-j \mathrm{sign}(\w).
\end{equation*}
This also means that the Fourier transform of $(\H f)(x)$ can then be expressed (using the convolution-multiplication rule) as
\begin{equation*}
\widehat {\H f} (\w)=-j \mathrm{sign}(\w) \hat f(\w).
\end{equation*}
Then by the Parseval-Plancherel identity, 
\begin{align*}
 \int | \widehat{\H f}(\w) |^2 d\w =  \int | \hat f(\w) |^2 d\w= \int |  f(x) |^2 dx
\end{align*}
for all $f$ in $\mathbf{S}$. This, in particular, establishes the fact that $\widehat {\H f}$, and hence $\H f$, is in $\eL^2$ for all $f \in \mathbf{S}$. In other words, $\H$ takes $\mathbf{S}$ into $\eL^2$, and that it is unitary:
\begin{align*}
\int | \H f(x) |^2 dx =  \int |f(x) |^2 dx    \qquad  (f \in \mathbf{S}).
\end{align*}
This establishes the estimate in \eqref{strong22}.

We can now extend the domain of $\H$ from the dense subclass $\mathbf{S}$ to $\eL^2$ using a continuity argument. For example, given an arbitrary function $f$ in $\eL^2$, we consider an approximating sequence $(f_n) \in \mathbf{S}$ such that $\|f_n-f\|_2$ can be made arbitrarily small for sufficiently large $n$. Then, using the \textit{strong} $(2,2)$ one can easily verify that the sequence $(\H f_n)$ is Cauchy in $\eL^2$. We define $\H f$ to be the limit of this Cauchy sequence (this is known to exist and is unique). This new operator, which we continue to denote by $\H$, is also bounded:
\begin{equation*}
\|\H f\|_2 = \|f\|_2 \qquad (\text{for all } f \in \eL^2).
\end{equation*}

\subsection{Weak $(1,1)$ nature}

As it turns out, the Hilbert transform is not bounded on $\eL^1$ and we would have to use a completely different set of tools to describe its behavior on this space. Before going through the details, we will first highlight the main difficulties involved in the derivation of the estimate and the strategies we use to handle them.

1. \textbf{Control of measures using norms}.  Let us first comment on the reason why the bound in \eqref{weak11} is termed as ``weak''.  Note that using the Chebyschev inequality, we can write
\begin{equation*}
|\{x: |f(x)| \geq \lambda\}| \leq \frac{\|f\|_1}{\lambda}
\end{equation*}
provided that $f$ is integrable (we use $|A|$ to denote the Lebesgue measure of some measurable subset $A$ of $\R$). In particular, if $T$ is a bounded operator on $\eL^1$, so that $\|Tf\|_1 \leq C\|f\|_1$ for all $f$ in $\eL^1$, we have
\begin{equation*}
|\{x: |Tf(x)| > \lambda\}| \leq \frac{\|Tf\|_1}{\lambda} <  C \frac{\|f\|_1}{\lambda}.
\end{equation*}
This tells us that  $T$ is also weakly bounded on $\eL^1$. The reverse assertion however is not true in general. For the Hilbert transform, one can easily see that $\H$ is not bounded on $\eL^1$ by considering the indicator function $\chi_{[0,1]}(x)$ and its Hilbert transform $\H \chi_{[0,1]}(x)$. An explicit computation shows that $\H \chi_{[0,1]}(x)$ decays only as $O(1/|x|)$ for large $x$ (besides having ``blow-ups'' at $0$ and $1$) and hence is clearly not integrable.

2. \textbf{Use of the weak bound}. Since we have already shown $\H$ to be bounded on $\eL^2$, one can however hope to salvage the situation at least for the intermediate $\eL^p$ spaces ($1 < p \leq 2$) by using an interpolation argument. This is exactly where the Marcinkiewicz interpolation theorem comes to the rescue, which roughly states that if $T$ is a weakly bounded linear operator on $\eL^p$ and $\eL^q$, then $T$ is strongly bounded on $\eL^r$ for all $p< r <q$. In particular, we will show that  $\H$  is weakly bounded on $\eL^1$, whereby the fact that  that $\H$ is bounded on $\eL^p$ for $1 < p \leq 2$ will be immediately established.

3. \textbf{Application on bounded integrable functions}. In order to derive the weak bound, it is clear from the above discussion (particularly one on the Chebyschev inequality) that one would be required to bound integrals of the form
\begin{equation*}
\int | \H f(x)|^n dx \qquad (n=1,2)
\end{equation*}
by the $\eL^1$ norm of $f$ (of course, assuming that $\H f(x)$ exists almost everywhere). It is however not clear whether this can be done for all Schwartz (or integrable) functions. This can be achieved under two distinctive situations.

The first among these is the case where the function $g(x)$ is both integrable and bounded; one can then verify that $g \in \eL^p$ for all $1\leq p \leq \infty$ (this itself is a kind of interpolation result). Indeed, using the fact that $\H$ is strongly bounded on $\eL^2$, we have for $n=2$,
\begin{equation}
\label{B1}
\int |\H g(x)|^2 dx = \int |g(x)|^2 dx \leq \|g\|_{\infty} \|g\|_1.
\end{equation}

4. \textbf{Application on localized oscillating functions}. The second case is the more interesting one which fundamentally relies on the odd nature of the kernel $1/x$. This is the case when the function under consideration $b(x)$ is well-localized and has a zero integral (oscillating).  Indeed, if the support of $b(x)$ is restricted to an interval $I$ and if $\int_I b =0$, then we can entirely avoid the limiting argument in \eqref{def} to write
\begin{align*}
\H b(x)= \frac{1}{\pi} \int_I \frac{b(y)}{x-y} dy= \frac{1}{\pi}\int_I b(y)\Big(\frac{1}{x-y}-\frac{1}{x-c}\Big) dy
\end{align*}
provided that  $x$ lies outside $I$, where $c$ denotes the centre of the interval. If we denote by $2I$ the interval having the same centre but twice the length as $I$,
then we see that
\begin{align*}
\int_{\R \backslash 2I} |\H b(x)| dx & \leq \frac{1}{\pi}\int_{\R \backslash 2I} \Big(\int_I |b(y)| \Big|\frac{1}{x-y}-\frac{1}{x-c}\Big|dy\ \Big)  dx \\
& = \frac{1}{\pi} \int_I |b(y)| \Big( \int_{\R \backslash 2I} \frac{|y-c|}{|x-y| |x-c|} dx \Big) dy \\
& < \frac{1}{\pi} \int_I |b(y)| \Big( \int_{\R \backslash 2I} \frac{|I|}{(x-c)^2} dx \Big) dy
\end{align*}
since $|y-c| < |I|/2$ for all $y \in I$, and $|x-y| > |x-c|/2$ for all $x$ outside $2I$. The inner integral is computed to be
\begin{equation*}
 \int_{\R \backslash 2I} \frac{|I|}{(x-c)^2} dx =2,
\end{equation*}
which provides us the the estimate
 \begin{equation}
\label{B2} 
\int_{\R \backslash 2I} |\H b(x)| dx < C \|b \|_1.
\end{equation}
We have however avoided a certain neighborhood of the support of $b(x)$ while evaluating the integral of $|\H b(x)|$. As will be seen shortly, this does not pose much of a problem since we can always control the size of this excluded interval by the norm of the function.

We are now in a position to derive \eqref{weak11}. We will do this only for non-negative functions; this will suffice since we can decompose any arbitrary function into its positive and negative parts, apply the result to each of them, and recombine the estimates. 

Following the above arguments, the main strategy would be to decompose the function $f(x)$ into a bounded and integrable part $g(x)$, and a series of localized oscillating bumps denoted by $b(x)$. The following version of a classic result of Calder\'on and Zygmund tells us that every integrable function (Schwartz functions in particular) can indeed be resolved in this manner (cf. Appendix A for details):

\vspace{5mm}

\textbf{The Calder\'o-Zygmund decomposition}.  Let $f$ be an non-negative integrable function on $\R$ and $\lambda$ be a positive number. Then there exists a sequence of almost disjoint intervals $\{I_k\}$ such that
\begin{description}
\item (i) $f(x) \leq \lambda$ for almost every $x$ outside $\Omega=\bigcup_k I_k$,
\item  (ii) The size of $\Omega$ is controlled by $f$, $|\Omega| \leq \lambda^{-1} ||f||_1$, and 
\item (iii) For every $I_k$, $\lambda < |I_k|^{-1} \int_{I_k} f(x) dx < 2\lambda$.
\end{description}

Let us set
\begin{equation*}
g(x)=
\begin{cases} f(x)  & \text{ for } x \in \Omega^c, \\
 |I_k|^{-1} \int_{I_k} f(x) dx & \text{ for } x \in I_k.
\end{cases}
\end{equation*}
It is clear that $g(x) \leq 2\lambda$ and $\| g\|_1=\|f\|_1$. We then set $b(x)=f(x)-g(x)$. This can be written as $b(x)=\sum_k b_k(x)$, where each $b_k(x)$ is defined to be
\begin{equation*}
b_k(x)=
\begin{cases} f(x)- |I_k|^{-1} \int_{I_k} f(x) dx & \text{if } x \in I_k, \\
 0 & \text{otherwise}.
\end{cases}
\end{equation*}
It is clear that $b_k$ is supported on the interval $I_k$ where $\int_{I_k} b_k(x) dx=0$, and that $\|b\|_1 \leq 2\|f\|_1$.

Since $\H f(x)=\H g(x) +\H b(x)$, 
\begin{align*}
|\{x: |\H f(x)| \geq \lambda\}| &\leq |\{x: |\H g(x)|  \geq \lambda/2\}| +|\{x: |\H b(x)|  \geq \lambda/2\}|.
\end{align*}
Using \eqref{B1} and the fact that $\| g\|_1=\|f\|_1$, we can use Chebyschev to get
\begin{equation}
\label{part1}
 |\{x: |\H g(x)| \geq \lambda/2\}| \leq \frac{A}{\lambda} \|f\|_1.
\end{equation}
To estimate $|\{x: |\H b(x)| > \lambda/2\}|$, we consider the union $\Omega^{\star}=\bigcup_k 2I_k$ of size $|\Omega^{\star}| \leq 2|\Omega|$. Then using estimate \eqref{B2} and the fact that\footnote{this is obvious if the sum is finite; there is some mild technicality involved in doing the same for an infinite sum.} $|\H b| \leq \sum_k |\H b_k|$ almost everywhere, we have
\begin{align}
\label{part2}
|\{x: |\H b(x)| \geq \lambda/2\}| & \leq |\Omega^{\star}| + |\{x \in \R \backslash \Omega^{\star}: |\H b(x)| \geq \lambda/2\}|  \nonumber \\
& \leq \frac{2}{\lambda} \|f\|_1 + \sum_k |\{x \in \R \backslash 2I_k: |\H b_k(x)| \geq \lambda/2\}|  \nonumber \\
& \leq \frac{2}{\lambda} \|f\|_1 + \frac{2}{\lambda}\int_{\R \backslash 2I_k} |\H b_k(x)| dx  \nonumber \\
& \leq \frac{B}{\lambda} \|f\|_1.
\end{align}
Combining \eqref{part1} and \eqref{part2}, we get
\begin{equation*}
|\{x: |\H f(x)|  \geq \lambda\}| \leq  \frac{C}{\lambda} \|f\|_1.
\end{equation*}

This establishes the desired \textit{weak} $(1,1)$ bound for the Hilbert transform. Based on an approximation argument, similar to the one used earlier for extending the domain of $\H$ from $\mathbf{S}$ to $\eL^2$ and using the notion of convergence in measure instead of norm, we can define the Hilbert transform $\H f$ of a function $f$ in $\eL^1$, which satisfies the estimate
\begin{equation*}
| \{x: |\H f(x)| \geq \lambda \}| \leq \frac{C}{\lambda} \|f\|_1.
\end{equation*}

\subsection{Strong $(p,p)$ nature using interpolation}

Since $\H$ is linear and we have shown that it is \textit{weak} $(p,p)$ for $p=1$ and $2$, we can conclude from the Marcenkiewicz interpolation theorem that $\H$ must be \textit{strong} $(p,p)$ for all $1 < p\leq 2$ (cf. Appendix B for details). This result can be extended to $2 < p < \infty$ using duality between the conjugate spaces $\eL^p$ and $\eL^q$ where $1/p+1/q=1$. The particular result we need is that
\begin{equation*}
\| f\|_q = \sup_{\|g\|_p=1} \Big | \int f(x) g(x) dx \ \Big|.
\end{equation*}

In particular, for $f \in \eL^q, \ 2 < q < \infty$, we can use this duality along with the fact that $\H$ is skew adjoint (this can be established for the Schwartz class and then extended using the usual density argument), namely that 
\begin{equation*}
\int (\H f)(x) g(x) =-\int f (x)(\H g)(x),
\end{equation*}
to write
\begin{align*}
\| \H f\|_q &= \sup_{\|g\|_p=1} \Big |\int \H f(x) g(x) dx \  \Big| \\
& = \sup_{\|g\|_p=1} \Big |\int \H g(x) f(x) dx  \ \Big| \\
& \leq \sup_{\|g\|_p=1} \| \H g\|_p \|f \|_q && (\text{Holders' inequality}) \\
& \leq C \|f \|_q. && (\H \text{ is bounded on } \eL^p)
\end{align*}
This establishes that $\H$ is \textit{strong} $(q,q)$ for $2 < q < \infty$. Note that the trick used in the above argument can be applied to any operator that is self-adjoint up to a sign, that is, to establish the boundedness of the operator on $\eL^p, 1 < p < \infty$, it suffices to do so only on the $\eL^p, 1 < p < 2$, or on $\eL^p, 2 < p < \infty$.

\section*{Appendix A}

To keep this note as self-contained as possible, we formulate the particular version of the Calder\'on-Zygmund decomposition used in establishing the weak $(1,1)$ bound, namely that if $f(x)$ is a non-negative integrable function and that $\lambda >0$, then there exists a sequence of almost disjoint (at most countable) intervals $\{I_k\}$ such that the following hold:
\begin{description}
\item (i) $f(x) \leq \lambda$ for almost every $x$ not belonging to $\bigcup_k I_k$;
\item (ii) The total length of the intervals $\{I_k\}$ is controlled by the norm of $f$,
\begin{equation}
\label{total_length}
\big |\bigcup_k I_k \big | \leq \frac{1}{\lambda} ||f||_1.  
\end{equation}
\item (iii) The average of $f$ on every $I_k$ is uniformly bounded,
\begin{equation}
\label{bounds}
\lambda < \frac{1}{|I_k|} \int_{I_k} f< 2 \lambda.
\end{equation}
\end{description}

This is the so-called Calder\'on-Zygmund decomposition of $f$ at height $\lambda$ and can be achieved using the following dyadic decomspoition strategy. We begin by partitioning $\R$ into a mesh of intervals, whose interiors are disjoint and whose common length is so large that  $|I|^{-1} \int_{I} f  \leq \lambda $ for every $I$ in this mesh (this can clearly be achieved since $|I|^{-1} \int_{I} f $ approached zero as $|I|$ gets large). 

Let $I_0$ be a fixed interval in this mesh. We split $I_0$ into two equal intervals. If we denote one of these intervals by $I_1$, then we have two distinct possibilities, namely that either
\begin{equation*}
 \frac{1}{|I_1|} \int_{I_1} f  > \lambda.
\end{equation*}
or
\begin{equation*}
\frac{1}{|I_1|} \int_{I_1} f  \leq \lambda.
\end{equation*}
In the former case, we do not split $I_1$ any further and $I_1$ is selected to be one of the intervals $I_k$ appearing in the decomposition.
We have for it \eqref{bounds}, because
\begin{equation*}
 \lambda < \frac{1}{|I_1|} \int_{I_1} f   \leq \frac{1}{2^{-1}|I_0|} \int_{I_0} f \leq 2 \lambda.
\end{equation*}

In the latter case, we split $I_1$ and repeat the process until we are forced into the former case (if this happens at all). We repeat this process starting with every interval from the initial mesh. Clearly, the resulting intervals $I_k$ are countable, and are almost disjoint by construction. To derive \eqref{total_length}, we note that
\begin{equation*}
\big |\bigcup_k I_k \big | = \sum_k |I_k| \leq \frac{1}{\lambda} \sum_k   \int_{I_k} f = \frac{1}{\lambda} \int_{\bigcup_k} f \leq \frac{1}{\lambda} \|f\|_1.
\end{equation*}

Finally, the fact that $f(x) \leq \lambda$ for almost every $x$ outside $\bigcup_k I_k$ can be deduced from the Lebesgue differentiation theorem, which states that the relation
\begin{equation}
\label{LebDiff}
f(x) = \lim_{|I| \rightarrow 0} \frac{1}{|I|} \int_I f(x-y) dy
\end{equation}
holds for almost every $x$ if $f$ is integrable (this is the Lebesgue counterpart of the fundamental theorem of calculus). Indeed, for every $x$ belonging to the complement $\bigcup_k I_k$ and for sufficiently small $I$ (this might not hold for certain larger intervals containing $x$; however this is inconsequential as we only need to consider the limiting case involving sufficiently small intervals), we have by construction 
\begin{equation*}
\frac{1}{|I|} \int_{I} f(x-y) dy  \leq \lambda.
\end{equation*}
Taking limits as $|I| \rightarrow 0$ and by applying \eqref{LebDiff} we see that $f(x) \leq \lambda$ for almost every $x$ outside $\bigcup_k I_k$. This establishes all the properties of the decomposition.

\section*{Appendix B}

	The \textit{weak} $(p,p)$ estimates for $\H$ for $p=1,2$ can be leveraged to a \textit{strong} $(p,p)$ estimate for $1 < p<2$ using interpolation. The first step in doing so is by breaking up a function $f \in \eL^p$, for some $1 < p<2$, as $f=g+h$, where $g$ and $h$ are in $\eL^1$ and $\eL^2$ respectively. One way of achieving this decomposition is by truncating $|f|$ on its range, that is, by setting $g=f \chi_{\{ |f| > \lambda\}}$ and $h=f \chi_{\{ |f|  \leq  \lambda\}}$ for some $\lambda >0$ (typically $h$ is the tail part of $f$ and $g$ is the singular part of $f$). Since $1-p<0$ and $2-p>0$, we see that
\begin{equation*}
\int |g| = \int_{|f| > \lambda} |f|^p |f|^{1-p} \leq \lambda^{1-p}  \int_{|f| > \lambda} |f|^p \leq \lambda^{1-p} \|f\|_p,
\end{equation*}
and
\begin{equation*}
\int |h|^2 = \int_{|f| \leq \lambda} |f|^p |f|^{2-p} \leq \lambda^{2-p}  \int_{|f| \leq \lambda} |f|^p < \lambda^{2-p} \|f\|_p,
\end{equation*}
so that $g$ and $h$ are indeed in the appropriate spaces.

To proceed further, we need the concept of a distribution function. Let us denote the number $|\{x: |f(x)| \geq \alpha\}|$, corresponding to some $\alpha >0$ and some function $f$, by $D_f(\alpha)$. The reason for introducing this function is that $D_f(\lambda)$ contains sufficient information for the evaluation of the $\eL^p$ norm of $f$. Indeed, by setting $E_{\alpha}=\{x: |f(x)| \geq \alpha\}$, we can write
\begin{align*}
 p \int_{0}^{\infty} \alpha^{p-1} D_f(\alpha) d \alpha &= p \int_{0}^{\infty} \alpha^{p-1} \Big( \int_{\R} \chi_{E_{\alpha}}(x) dx \Big) d \alpha  \\
 &=  \int_{\R} \Big( \int_{0}^{\infty} p \alpha^{p-1} \chi_{E_{\alpha}}(x) d\alpha \Big) dx  \\
&=  \int_{\R} \Big( \int_{0}^{|f(x)|} p \alpha^{p-1} d\alpha \Big) dx \\
& = \int_{\R} |f(x)|^p dx ,
\end{align*}
which gives us the equivalence
\begin{equation*}
\|f\|_p^p = p \int_{0}^{\infty} \alpha^{p-1} D_f(\alpha) d \alpha.
\end{equation*}

Now, using the \textit{weak} $(p,p)$ estimates ($p=1,2$), we have 
\begin{equation}
\label{est1}
D_{\H g}(\lambda)  \leq \frac{C}{\lambda} \| g \|_1 = \frac{C}{\lambda} \int_{|f| > \lambda} |f|.
\end{equation}
Similarly,
\begin{equation}
\label{est2}
D_{\H h}(\lambda)  \leq \frac{1}{\lambda^2} \| h \|^2_2 \leq  \frac{1}{\lambda^2}  \int_{|f| \leq \lambda} |f|^2. 
\end{equation}

The rest of the computation is based on the properties of $D_f(\alpha)$ and the interplay of the indices of the $\eL^p$ spaces. In particular, since $\H f (x)= \H g(x) + \H h(x)$, one can verify that $D_{\H f}(\lambda) \leq D_{\H g}(\lambda/2)+ D_{\H h}(\lambda/2)$. By combining this with \eqref{est1} and \eqref{est2}, we have
\begin{align*}
\|\H f\|_p^p &= p \int_{0}^{\infty} \alpha^{p-1} D_{\H f}(\alpha) d \alpha \\
& \leq p \int_{0}^{\infty} \alpha^{p-1} D_{\H g}(\alpha/2) d \alpha + p \int_{0}^{\infty} \alpha^{p-1} D_{\H h}(\alpha/2) d \alpha \\
& \leq B_1 \int_{0}^{\infty} \alpha^{p-2}   \Big(\int_{|f| > \lambda} |f| \ \Big) d\alpha + B_2 \int_{0}^{\infty} \alpha^{p-3}   \Big(\int_{|f| \leq \lambda} |f|^2 \ \Big) d\alpha \\
& \leq  B_1 \int_{\R} \Big ( \int_0^{|f|} \alpha^{p-2} d\alpha \ \Big)  |f|  +  B_2 \int_{\R} \Big ( \int_{|f|}^{\infty} \alpha^{p-3} d\alpha \ \Big)  |f|^2 \\
& = C \|f\|^p_p.
\end{align*}
This shows that $\H$ is bounded on $\eL^p$ for $1 <p<2$.

\begin{remark} The above was a special case of the so-called Marcenkiewicz interpolation theorem. The above derivation can easily be extended to the most general form of the theorem which states that if $T$ is sub-linear (i.e., $|T(f+g)| \leq |Tf| + |Tg|$) and is both \textit{weak} $(p,p)$ and \textit{weak} $(q,q)$, then $T$ is \textit{strong} $(r,r)$ for all $1 
\leq p<r< q \leq \infty$. The fact that the weaker hypothesis of sub-linearity would suffice is also clear from the above derivation since the sub-additivity property $D_{T(f+g)}(\alpha) \leq D_{Tf}(\alpha)+D_{Tg}(\alpha)$ holds even when $T$ is sub-linear.
\end{remark}

\bibliographystyle{amsplain}
\bibliography{HT_SingularIntegrals.bib}

\providecommand{\bysame}{\leavevmode\hbox to3em{\hrulefill}\thinspace}
\providecommand{\MR}{\relax\ifhmode\unskip\space\fi MR }
\providecommand{\MRhref}[2]{%
  \href{http://www.ams.org/mathscinet-getitem?mr=#1}{#2}
}
\providecommand{\href}[2]{#2}
\begin{thebibliography}{1}

\bibitem{Grafakos}
L.~Grafakos, \emph{Classical and {M}odern {F}ourier {A}nalysis}, Prentice Hall,
  2003.

\bibitem{Riesz}
M.~Riesz, \emph{Sur les fonctions conjugu\'ees}, Mathematische {Z}eitschrift
  (1928), 218--244.

\bibitem{Stein}
E.~M. Stein, \emph{Singular {I}ntegrals and {D}ifferentiability {P}roperty of
  {F}unctions}, Princeton University Press, 1970.

\end{thebibliography}

\end{document}